\documentclass[times, twoside]{Preprint}

\usepackage{amsmath}
\usepackage{multirow}
\usepackage{blindtext}
\usepackage{graphicx}
\usepackage{dblfloatfix}
\usepackage{color, colortbl}
\usepackage{url}
\usepackage{upgreek}
\usepackage[breakable]{tcolorbox}

\definecolor{lightgray}{rgb}{0.95, 0.95, 0.95}
\definecolor{salmon1}{rgb}{1.0, 0.63, 0.48}
\definecolor{salmon2}{rgb}{1.0, 0.87, 0.68}
\definecolor{salmon3}{rgb}{1.0, 0.94, 0.84}
\definecolor{gray1}{rgb}{0.7, 0.7, 0.7}
\definecolor{gray2}{rgb}{0.85, 0.85, 0.85}
\definecolor{blue1}{rgb}{0.69, 0.77, 0.87}
\definecolor{blue2}{rgb}{0.94, 0.97, 1.0}
\definecolor{blue3}{rgb}{0.88, 1.0, 1.0}

\leadauthor{M. Ballester} 

\begin{document}

\setlength{\parindent}{1cm}

\title{Characterization of Crystal Properties and Defects in \\CdZnTe Radiation Detectors}

\shorttitle{Crystal characterization in CZT detectors}

\setlength{\affilsep}{7 pt}
\renewcommand\Authfont{\color{black}\normalfont\sffamily\bfseries\fontsize{9}{11}\selectfont}
\renewcommand\Affilfont{\color{black}\normalfont\sffamily\fontsize{6.5}{8}\selectfont}
\renewcommand\Authands{, and }

\author[1]{Manuel Ballester}
\author[2]{Jaromir Kaspar}
\author[2]{Francesc Massan\'es}
\author[3]{Srutarshi Banerjee}
\author[2]{\\Alexander Hans Vija}
\author[1, 4]{Aggelos K. Katsaggelos}

\affil[1]{Department of Computer Sciences, Northwestern University, Evanston, IL 60208, USA}
\affil[2]{Siemens Medical Solutions USA Inc., Hoffman Estates, IL 60192, USA}
\affil[3]{X-Ray Science Division, Argonne National Laboratory, Lemont, IL 60439, USA}
\affil[4]{Department of Electrical and Computer Engineering, Northwestern University, Evanston, IL 60208, USA}

\affil[**]{Correspondence: manuelballestermatito2021@u.northwestern.edu}

\maketitle

\begin{abstract}

CdZnTe-based detectors are highly valued because of their high spectral resolution, which is an essential feature for nuclear medical imaging. However, this resolution is compromised when there are substantial defects in the CdZnTe crystals. In this study, we present a learning-based approach to determine the spatially dependent bulk properties and defects in semiconductor detectors. This characterization allows us to mitigate and compensate for the undesired effects caused by crystal impurities. We tested our model with computer-generated noise-free input data, where it showed excellent accuracy, achieving an average RMSE of 0.43\% between the predicted and the ground truth crystal properties. In addition, a sensitivity analysis was performed to determine the effect of noisy data on the accuracy of the model.

\end {abstract}

\vspace{-3mm}

\section{Introduction}
Single-photon detectors are essential in nuclear medical imaging techniques, where high spectral resolution is critical for tissue differentiation. In particular, detector technologies based on cadmium zinc telluride (CdZnTe) are increasingly favored in modalities that operate with relatively low photon flux, such as \textit{single photon emission computed tomography} (SPECT) \cite{iniewski2014czt}. CdZnTe crystals exhibit a high stopping power (allowing the absorption of most incoming photons) and a wide band gap of 1.6 eV (allowing detector operation at room temperature) \cite{chun2008property}. These photon-counting detector (PCDs) offer a rapid temporal response (less than 10 ns) and can be configured with a pixelated geometry, enabling submillimeter spatial resolution. However, the energy resolution of these detectors is often compromised due to the high defect density typically found in the thick CdZnTe crystals \cite{defects1, defects2, defects3}. These defects include point defects, dislocations, and grain boundaries \cite{types_defects}, which adversely influence the optoelectric properties of the crystal.

\begin{figure}[h!]
\centering
\includegraphics[width=0.95 \linewidth]{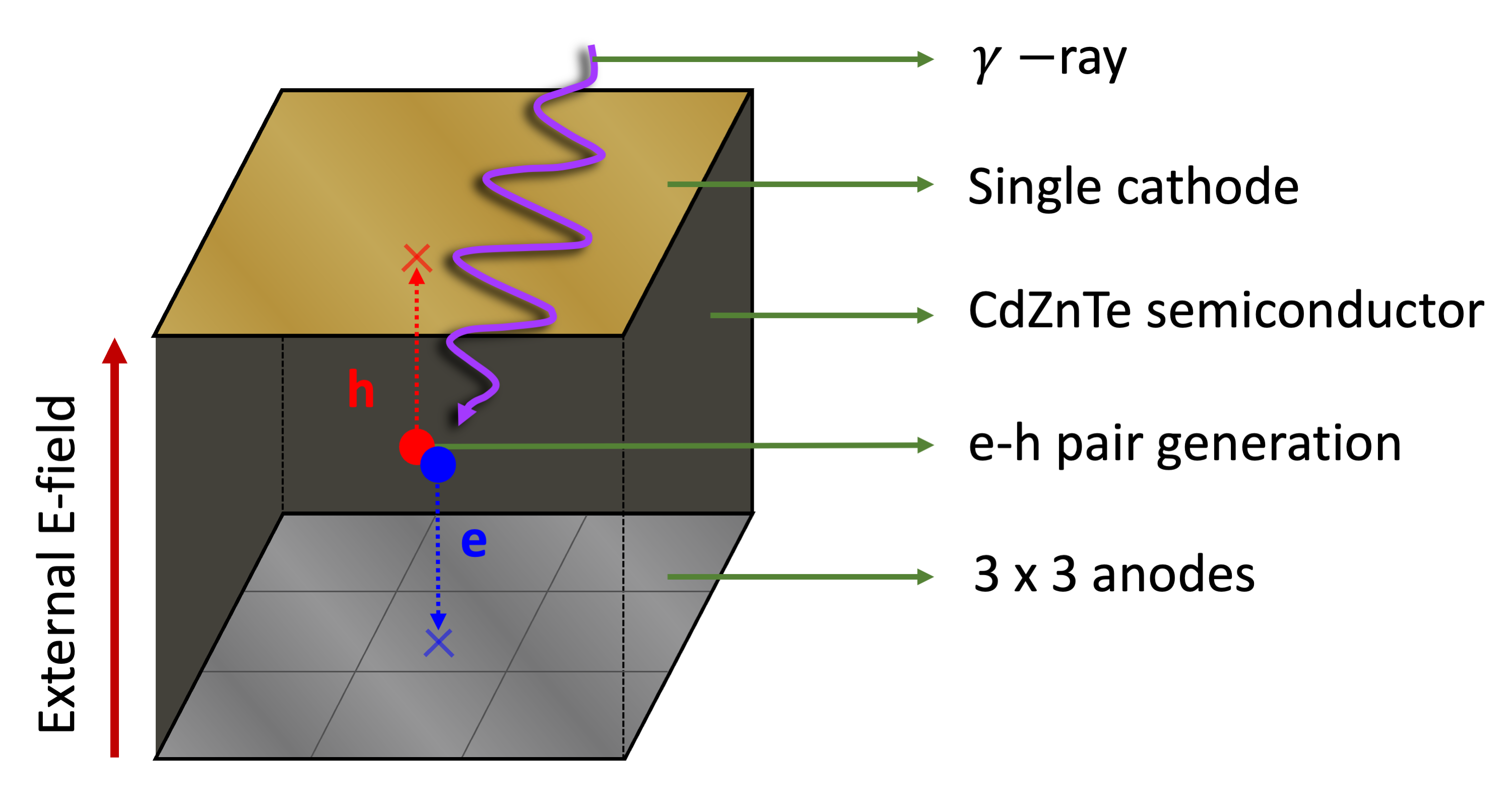}
\caption{Pixelated detector geometry with one cathode on the top and nine anodes on the bottom. }
\label{fig:detector}
\end{figure}

In this study, we employ a reverse engineering approach to determine the material properties and defects in the CdZnTe crystals used for radiation detectors. Building upon prior physics-based learning models developed by our team \cite{ballester2022materials, table_Nature, banerjee2023machine, banerjee2023identifying, banerjee2024physics}, our novel model accurately identifies fundamental spatially-dependent properties, such as charge mobility and the lifetimes for charge recombination, trapping, and detrapping. By precisely characterizing these features, the study seeks to mitigate the detrimental effects of crystal impurities. The model takes as input the charge-induced signals and the charge concentrations after a particular photon-detector event interaction. We initially validate our model using computer-generated noise-free data to establish a baseline for performance. Subsequently, we now assess for the first time the impact of noise in the input data on the accuracy of this characterization.

\vspace{-0.2cm}
\section{Detector simulator}

\begin{figure*}[h!]
\centering
\includegraphics[width=0.95 \textwidth]{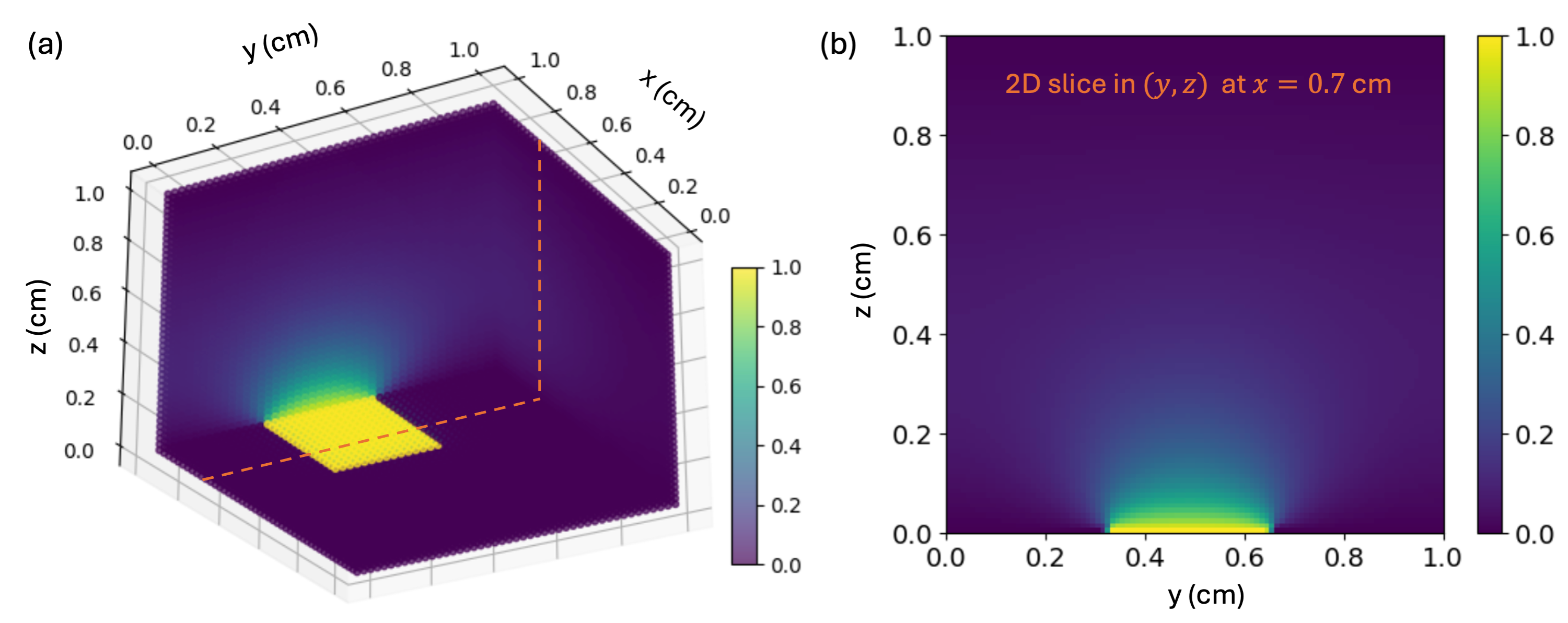}
\caption{Weighting potential for anode $k$ at the edge of the pixel array grid. (a) 3D view of the weighting potential. (b) 2D slice of the weighting potential at $x = 0.70$ cm.}
\label{fig:field}
\end{figure*}

\begin{table*}[b]
  \centering
  \caption{Common material properties of CdZnTe crystals \cite{ballester2024modeling}, along with the corresponding dimensionless computational parameters used in the simulations. The table also shows the NRMSE between the predicted parameters and the ground-truth ones when employing noise-free and noisy input data.}
  
  \begin{tabular}{|c|c|c|c|c|c|c|}
    \hline
    \multirow{2}{*}{Material properties} & \multirow{2}{*}{Symbol} & \multirow{2}{*}{Value} & \multirow{2}{*}{Parameter} & \multirow{2}{*}{Value} & \multicolumn{2}{c|}{NRMSE (\%)} \\
    \cline{6-7}
    & & & & & Noise-free & Noisy \\
    \hline
    \multirow{2}{*}{Charge mobility [cm$^2$/Vs]} & $\mu_\text{e}$ & 1120 & $R_\text{e}=(\mu_\text{e} E) \Delta t /\Delta z$ & 0.95 & $<10^{-2}$ & 0.22 \\
    \cline{2-7}
    & $\mu_\text{h} \approx \mu_\text{e}/10$ & 112 & $R_\text{h} = (\mu_\text{h} E) \Delta t /\Delta z$ & 0.095 & $<10^{-2}$ & 0.22 \\
    \hline 
    \multirow{2}{*}{Recombination lifetime [$\mu$s]} & $\tau_\text{eR}$ & 10 & $P_\text{eR}=\Delta t/\tau_\text{eR}$ & 0.001 & 2.81 & 23.81 \\
    \cline{2-7}
    & $\tau_\text{hR}$ & 1 & $P_\text{hR}=\Delta t/\tau_\text{hR}$ & 0.01 & $<10^{-2}$ & 2.86 \\
    \hline 
    \multirow{2}{*}{Trapping lifetime [$\mu$s]} & $\tau_\text{eT}$ & 10 & $P_\text{eT} =\Delta t/\tau_\text{eT}$ & 0.001 & 0.18 & 3.46 \\
    \cline{2-7}
    & $\tau_\text{hT}$ & 0.067 & $P_\text{hT} =\Delta t/\tau_\text{hT}$ & 0.15 & $<10^{-2}$ & 0.74 \\
    \hline 
    \multirow{2}{*}{Detrapping lifetime [$\mu$s]} & $\tau_\text{eD}$ & 0.4 & $P_\text{eD} = \Delta t/\tau_\text{eD}$ & 0.025 & 0.01 & 4.52 \\
    \cline{2-7}
    & $\tau_\text{hD}$ & 0.067 & $P_\text{hD} = \Delta t/\tau_\text{hD}$ & 0.15 & $<10^{-2}$ & 0.88 \\
    \hline
  \end{tabular}
  \label{tab:1}
\end{table*}

We have built a physics-driven model that simulates the functioning of the CdZnTe photon-counting detectors. Our simulations consider a detector with dimensions $1 \times 1 \times 1$ cm$^3$ and a standard pixelated configuration: a single cathode on top and nine anodes on the bottom (see Fig. \ref{fig:detector}). When a $\gamma-$ray penetrates the cathode and interacts with the crystal, several electron-hole (e-h) pairs are generated. Drifted by an external electric field $E$, these charge carriers move and produce charge-induced signals at the nearby electrodes \cite{SR_review, miesher_model}.

\noindent Following a delta-like photon-detector event $\delta(x_0, y_0, z_0, t_0)$ occurring at time $t_0$ and specific location $(x_0, y_0, z_0) \in [0,1]^3$, the electron concentration $n_\text{e}(x,y,z,t)$ can be model with a system of partial differential equations (PDEs) \cite{Th_trapping_detrapping, kamieniecki2014effect}:
\begin{equation}
 \begin{cases} 
 \partial_t n_\text{e} - \nabla\cdot (\mu_\text{e} E n_\text{e}) = - \frac{1}{\tau_\text{eR}} n_\text{e} - \frac{1}{\tau_\text{Te}} n_\text{e} + \frac{1}{\tau_\text{eD}} \tilde{n}_\text{e} + \delta\\
 \partial_t \tilde{n}_\text{e} = \frac{1}{\tau_\text{eT}} n_\text{e} - \frac{1}{\tau_\text{eD}} \tilde{n}_\text{e}\ 
 \end{cases} 
\label{eq:pde} 
\end{equation}
Note that an equivalent system and notation can also be applied to model the dynamics of holes, where the concentration is denoted as $n_\text{h}$. Eq. \ref{eq:pde} account for various dynamic processes: charge drift, charge generation-recombination, and trapping-detrapping effects. The variable $\tilde{n}_\text{e}$ represents the concentration of electrons within the trapping energy levels, which includes shallow and deep traps. These traps are conceptualized as an infinite \textit{wells} that attracts (and eventually expels) a percentage of the charges, as described by the Shockley-Read-Hall theory \cite{zimmerman1973experimental}.

The mean lifetimes for electron charge trapping, detrapping, and recombination are denoted by $\tau_\text{eT}$, $\tau_\text{eD}$ and $\tau_\text{eR}$, respectively. The electron mobility is given by $\mu_\text{e}$, with its drift velocity being $v_\text{e} = - \mu_\text{e} E$. In the simulations, we considered a uniform vertical field with magnitude $E = 850$ V/cm. Our model simplifies the charge dynamics by assuming a purely vertical trajectory for the charges along the $z-$axis, omitting minor deviations that could arise from interpixel gaps \cite{buttacavoli2022incomplete} or polarization effects \cite{bale2008nature}. Additionally, we incorporate an empirical relationship between the mobilities of electrons and holes, estimated as $\mu_\text{h} \approx \mu_\text{e}/10$, based on the findings reported in \cite{Srutarshi_IEEE}. Our model neglects the weak effects of diffusion and Coulomb repulsion, which is reasonable under a high electric field strength \cite{Th_trapping_detrapping, diffusion_coulomb,simulation_probabilistic}. 

Eq. \ref{eq:pde} can be efficiently solved numerically using the explicit finite-difference method \cite{ballester2024modeling}. We defined a spatial step of $\Delta z = 0.01$ cm and a time step of $\Delta t = 10$ ns, capturing the high temporal response of the real CdZnTe-PCDs. Note that the crystal properties (charge mobility and lifetimes) can be equivalently reformulated using dimensionless computational parameters, as detailed in Table \ref{tab:1}. These parameters naturally emerge during the discretization of the PDE system, as further explained in \cite{table_Nature}. To accommodate the spatial variability of properties, we define a stratified media: There are $N=100$ layers (stacked sequentially from the cathode to the anode), each with distinct parameter values. 

After numerically solving the system from Eq. \ref{eq:pde} for both electron and hole concentrations, we obtain the charge densities as functions of time and space. The current signals for each electrode, indexed by $k \in \{1, 2, ..., 10\}$, are subsequently calculated using the Shockley-Ramo theorem \cite{SR_review}:

\begin{equation}
    i^{(k)}(t) = q n_\text{e}(t,x,y,z) E^{(k)}_\text{w}(x,y,z),
\end{equation}
where the weighting electric field $E^{(k)}_\text{w}$ is derived from the corresponding weighting potential $\phi^{(k)}_\text{w}$. The weighting potential is determined by solving the Poisson equation $\nabla^2 \phi^{(k)}_\text{w}(x,y,z) = 0$, subject to the boundary conditions $\phi^{(k)}_\text{w} = 1$ at the location of the electrode $k$, and $\phi^{(k)}_\text{w} = 0$ at the areas of all other electrodes. Figure \ref{fig:field} illustrates the weighting potential for an anode located at the edge of the bottom pixel array. It is worth noting that the commonly measured charge-induced signals are simply calculated as $Q^{(k)}(t) = \int i^{(k)}(t) dt$.

\vspace{-0.2cm}
\section{Detector characterization}

\begin{figure*}[t!]
\centering
\includegraphics[width=0.95 \textwidth]{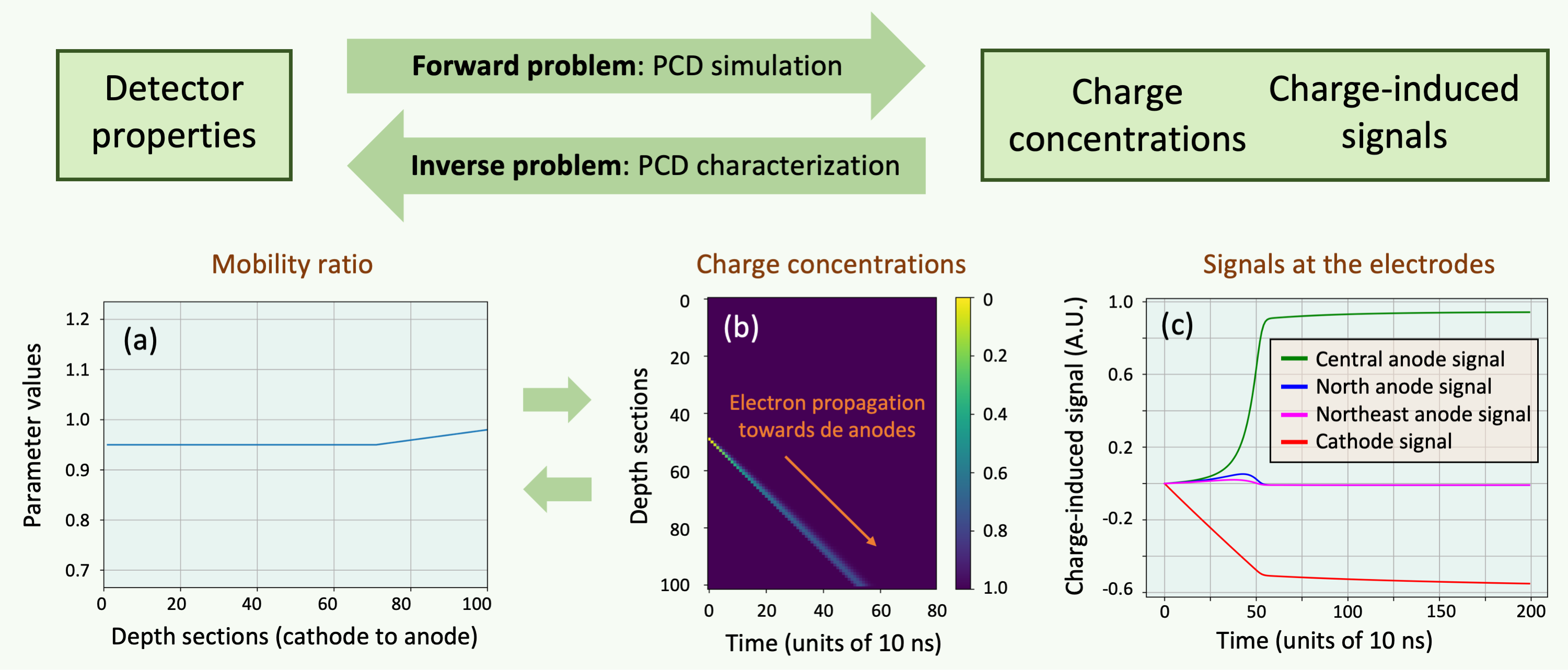}
\caption{Diagram illustrating the forward and inverse problems. In the forward problem (a), the model serves as a digital twin of the photon-counting detector, providing an accurate simulation of the detector signals based on the crystal properties, such as charge mobility ($\mu_\text{e}$), or equivalently, the computational parameter $R_\text{e}$ (mobility ratio). Conversely, the inverse problem (b) aims to deduce the material properties from the available information on the charges and resulting signals.}
\label{fig:forward_vs_inverse}
\end{figure*}

Given the detector signals and charge concentrations, we can now construct an inverse model that provides us with the material properties and defects of the CdZnTe crystal at different locations, as outlined in Fig. \ref{fig:forward_vs_inverse}. As proposed in \cite{table_Nature}, we aim to solve the following optimization problem:
\begin{equation}
  \begin{aligned}
    \min C(\Theta) = & \, \| n^\text{sim}_\text{e}(t,z;\Theta) - n^\text{given}_\text{e}(t,z) \|_\text{F}^2 \\
    & + \| n^\text{sim}_\text{h}(t,z;\Theta) - n^\text{given}_\text{h}(t,z) \|_\text{F}^2 \\
    & + \| \tilde{n}^\text{sim}_\text{e}(t,z;\Theta) - \tilde{n}^\text{given}_\text{e}(t,z) \|_\text{F}^2 \\
    & + \| \tilde{n}^\text{sim}_\text{h}(t,z;\Theta) - \tilde{n}^\text{given}_\text{h}(t,z) \|_\text{F}^2 \\
    & + \lambda \sum_{i=1}^{10} \| Q_i^\text{sim}(t;\Theta) - Q_i^\text{given}(t) \|_\text{F}^2
  \end{aligned}
  \label{eq:optimization}
\end{equation}

The 2D array $n^\text{sim}_\text{e}(t,z; \Theta)$ depicts the simulated concentration of electrons at discrete times and positions, and $\| \cdot\|_\text{F}$ the Frobenius matrix norm. The simulated concentration and signals depend on the computational parameters, which are encapsulated by $\Theta \in \mathbb{R}^{7 \times 100}$, seven parameters over the $N=100$ voxels in depth. Equation \ref{eq:optimization} describes a problem in which we try to fit the simulator's output to the provided data. It is crucial to note that the data used in this study were generated in a computational manner, not from experiments. This means that we have prior knowledge of the ground-truth parameters, enabling a straightforward evaluation of our inverse solver algorithm. The errors between the calculated and given charge-induced signals $Q^{(k)}$ at each electrode $k \in \{1,2,...,10\}$ are incorporated into the Eq. \ref{eq:optimization} as a regularization term with a relatively low coefficient ($\lambda = 0.1$), as proposed in \cite{table_Nature}.

The non-linear fitting problem from Eq. \ref{eq:optimization} displays a cost function $C(\Theta)$ with several local minima, and therefore describes a complicated global optimization problem. Due to the curse of dimensionality \cite{aggelos}, conventional global optimizers have a relatively high computational cost and are less suitable for our problem. To address this challenge, we employ the Adam optimizer \cite{DL_optimizers}, a momentum-based gradient descent method designed to handle non-convex optimization problems efficiently while avoiding stagnation at local minima. For that optimizer, we set the hyperparameters with a learning rate of $5 \cdot 10^{-4}$, a first moment $\beta_1 = 0.9$, and a second moment $\beta_2 = 0.999$. 

Please observe that we used automatic differentiation to accelerate gradient computations with the \textit{PyTorch} library. We also utilized an NVIDIA Tesla GPU to run our program, which is specifically designed to accelerate AI operations. We are able to perform 20,000 iterations, until convergence, in approximately 3 hours.

To evaluate the model outcomes, we adopt the Normalized Root Mean Square Error (NRMSE) of each parameter within a specified Region of Interest (ROI). For an event at the center of the detector (depth voxel 50), the ROI for the electron properties spans voxels 50-100, as most of the generated free electrons reach the anode location (voxel 100). In contrast, due to the relative lower drift velocity of holes, they predominantly undergo recombination or trapping when traveling through the first 10 voxels, being their ROI between voxels 40-50. Finally, it is important to note that we enforce box constraints using the projected gradient method \cite{boyd_convex}. The NRMSE could range $\pm$25\% relative to the values of the ground-truth parameters.

\vspace{-0.2cm}
\section{Results}
\begin{figure*}[h!]
\centering
\includegraphics[width=0.95 \textwidth]{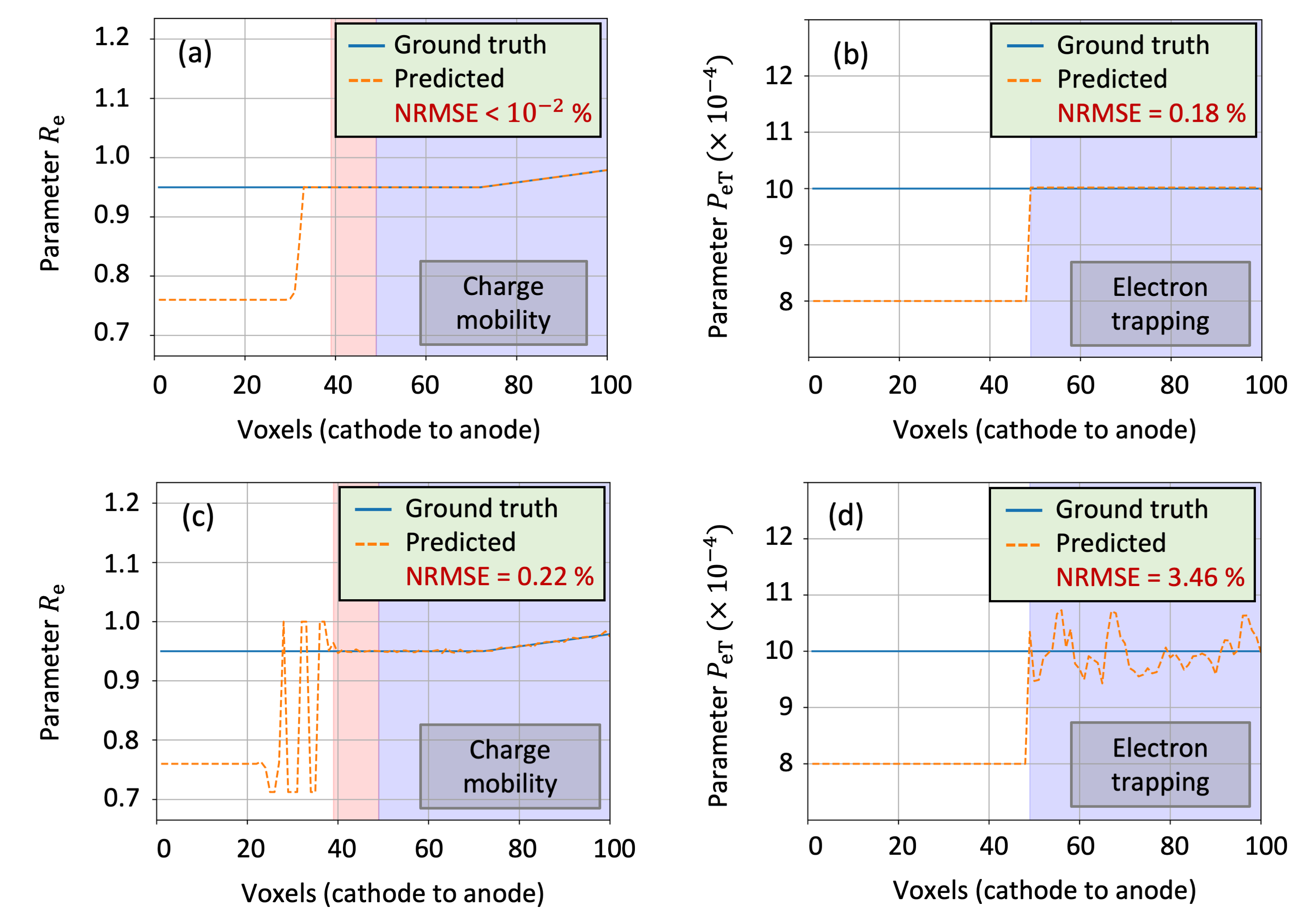}
\caption{Comparison of ground-truth and predicted parameters for (a,b) noise-free, and (c,d) noisy input data. (a,c) Mobility ratio, and (b,d) electron trapping parameter.}
\label{fig:parameters}
\end{figure*}

We first want to extract detector characteristics and defects considering ideally noise-free computer-generated input data. Figure \ref{fig:parameters} a shows both the predicted and ground-truth computational parameter $R_\text{e}$ (see Table \ref{tab:1}) in each depth voxel, indicative of charge mobility. Consequently, we are using a single mobility parameter for both electrons (with ROI highlighted in blue) and holes (ROI in red). Figure \ref{fig:parameters}b offers a fitting result for another representative parameter (associated with the electron trapping lifetime). Please find in Table \ref{tab:1} all the NRMSE found for the multiple computational parameters. The average error was only 0.43\% for all the parameters, indicating the correct convergence.

We will now analyze the case in which an additive Gaussian noise is introduced to the computer-generated input datasets. The standard deviation of the Gaussian noise is 0.1\% of the maximum signal value. It is crucial to note that the cost function $C$ should not reach zero in these noisy scenarios: A cost below the threshold given by the ground truth parameters, $C(\Theta^\text{gt})$, would indicate an overfit, suggesting that the model has learned the noise. The regularizer in Eq. \ref{eq:optimization} prevents such overfitting, as it inhibits the algorithm from adjusting both the noise in the concentrations and the signals simultaneously. Figures \ref{fig:parameters}c-\ref{fig:parameters}d show a comparison between the predicted parameters $R_\text{e}$ and $P_\text{eT}$ and their ground-truth values within the regions of interest for this case with noisy data. The average NRMSE of the parameters determined became 4.89\%. Although $R_\text{e}$ maintained a relatively low error of 0.22\%, the error for $P_\text{eT}$ increased to 3.46\% (see Fig. \ref{fig:parameters}). The last column of the Table \ref{tab:1} presents the results for all other parameters, in this case of noisy input data. In particular, we can see that the electron recombination parameter, $P_\text{eR}$, exhibits a significantly larger error. Its precise prediction during training becomes less critical because small variations in this parameter do not significantly impact the overall concentrations and signals \cite{ballester2024modeling}.

\begin{figure*}[h!]
\centering
\includegraphics[width=0.95 \textwidth]{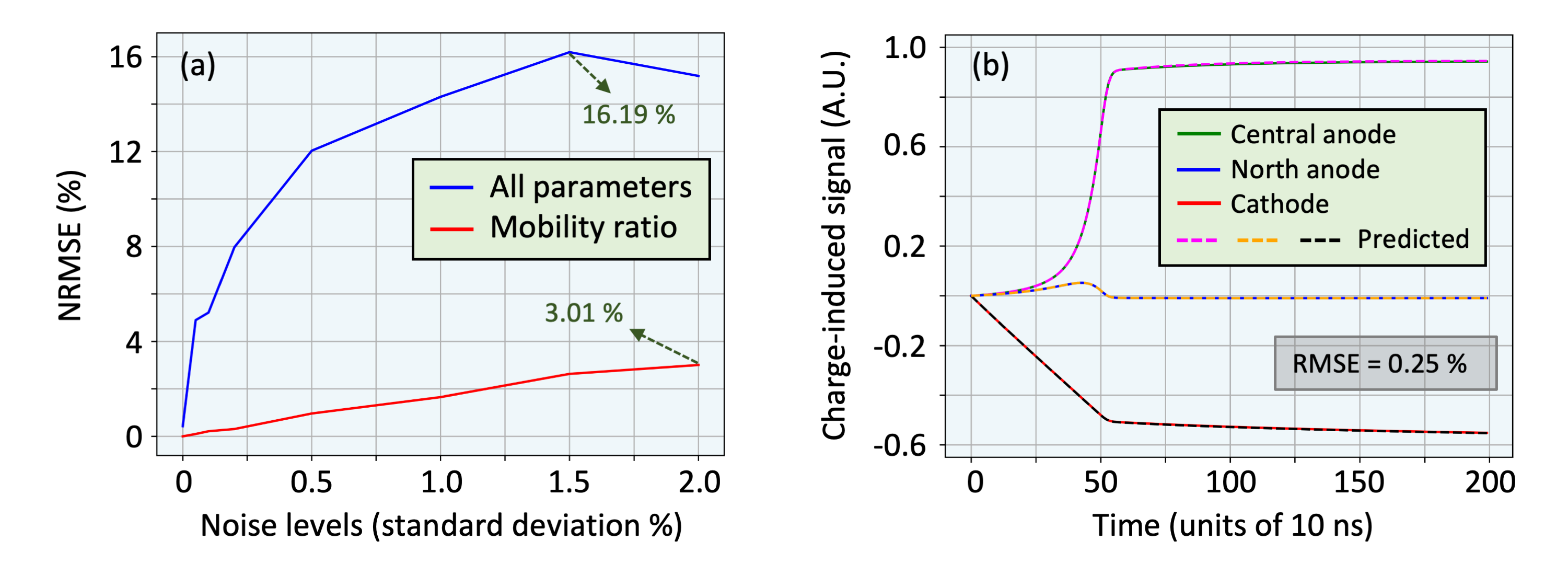}
\caption{(a) Average error in predicted parameters with varying noise levels in the input data. (b) Ground-truth signal compared to signals from predicted parameters under high noise conditions (std = 2\%).}
\label{fig:noise}
\end{figure*}

Figure \ref{fig:noise}a indicates the correlation between the increase in the standard deviation in the input data and the increase in the average error in the predicted parameters. With a standard deviation ranging from 0\% to 2\%, the highest observed NRMSE was 16.19\%. Interestingly, the random nature of the added noise and the ADAM optimizer itself led to the peak at 1.5\% and not 2.0\%, as one might initially anticipate. In contrast, the NRMSE for the mobility ratio increased almost linearly with the added noise level and found a maximum value of only 3.01\%. In Fig. \ref{fig:noise}b, one sees the ideal ground-truth signals compared to the signals generated by the predicted parameters. Even with an input noise level that has a standard deviation of 2\%, the predicted signals align closely with the ground truth, resulting in an error of 0.25\%.

\vspace{-0.2cm}
\section{Discussions}
Although our inverse model demonstrated high accuracy with noise-free input data, the introduction of additive Gaussian noise to the charge concentrations and signals led to more pronounced errors. It is important to emphasize that the model still yields valuable insights even under high noise levels, as explained in the next two arguments.

First, the initial parameters deviate by 20\% from the ground truth values, and the limit of the box constraint was up to $\pm25\%$. In contrast, the average error of our model remained well below 16.2\% for all scenarios (see Fig. \ref{fig:noise}a). These results demonstrate that even under pathologically noisy conditions, the algorithm still refines the initial estimates.

Second, throughout all noise conditions, the mobility ratio maintained a relatively low error. Slight variations in this parameter induce large changes in the signal production, and therefore in the cost function value \cite{ballester2024modeling}. Therefore, this high sensitivity enables the correct characterization of the charge mobility.

However, our approach is not without limitations. On the one hand, we recognize that our approach involves the so-called ``inverse crime'', since we employ the same simulator for the generation of input data (with or without noise) and to solve the optimization problem. Future studies should validate the model using real-world experimental data. On the other hand, it must be noted that, while our detector simulator \cite{ballester2024modeling} is efficient, it operates under simplified assumptions. As mentioned, it considers a uniformly directed external E-field from the anodes to the cathode (resulting in one-dimensional charge trajectories, in the $z-$ direction) and also neglects the charge cloud expansion in the 3D space due to the second-order effects of charge diffusion and Coulomb repulsion.

\vspace{-0.2cm}
\section{Conclusions}
This study introduces a reverse engineering model to infer spatially varying material properties and defects in CdZnTe detectors using charge concentrations and signals. Our developed software shows a relatively high accuracy (with an average error of 0.43\%) when using noise-free input data. However, the error increased to 4.89\% when processing data inputs affected by the additive Gaussian noise, with a standard deviation of 0.1\% of the signal peak. Despite the performance degradation with higher noise levels, the derived parameters still provide a refined understanding compared to the initial assumptions. Furthermore, our model accurately predicted the mobility ratio even under conditions of very high noise levels. Finally, it is important to highlight that the parameters related to the lifetimes of recombination, trapping, and detrapping are highly susceptible to noise disturbances in the input data.

\vspace{0.5 cm}
\noindent \textbf{Funding:} This work was supported by Siemens Medical Solutions USA, Inc. within the scope of a research agreement. 

\noindent \textbf{Disclosures:} We disclose that F.M., J.K. and A.H.V. are currently employed by Siemens Healthineers.

\noindent \textbf{Data Availability Statement:} Data underlying the results presented in this article are not publicly available at this time, but may be obtained from the authors upon reasonable request.

\noindent \textbf{Acknowledgments:} The authors thank Wendy Mcdougald and Ron Malmin for reviewing the manuscript. The authors thank Miesher Rodrigues for his assistance.

\section*{Bibliography}
\bibliography{refs}

\begin{thebibliography}{25}
\providecommand{\natexlab}[1]{#1}
\providecommand{\url}[1]{\texttt{#1}}
\expandafter\ifx\csname urlstyle\endcsname\relax
  \providecommand{\doi}[1]{doi: #1}\else
  \providecommand{\doi}{doi: \begingroup \urlstyle{rm}\Url}\fi

\bibitem[Iniewski(2014)]{iniewski2014czt}
K~Iniewski.
\newblock Czt detector technology for medical imaging.
\newblock \emph{Journal of Instrumentation}, 9\penalty0 (11):\penalty0 C11001, 2014.

\bibitem[Chun et~al.(2008)Chun, Park, Lee, Kim, Ha, Kang, Cho, Hong, and Kim]{chun2008property}
Sung-Dae Chun, Se-Hwan Park, Dong~Hoon Lee, Yong~Kyun Kim, Jang~Ho Ha, Sang~Mook Kang, Yun~Ho Cho, Duk-Geun Hong, and Jong~Kyung Kim.
\newblock Property of a czt semiconductor detector for radionuclide identification.
\newblock \emph{Journal of Nuclear Science and Technology}, 45\penalty0 (sup5):\penalty0 421--424, 2008.

\bibitem[Bolotnikov et~al.(2009)Bolotnikov, Babalola, Camarda, Chen, Awadalla, Cui, Egarievwe, Fochuk, Hawrami, Hossain, et~al.]{defects1}
Aleksey~E Bolotnikov, Stephen~O Babalola, Giuseppe~S Camarda, Henry Chen, S~Awadalla, Yonggang Cui, Stephrn~U Egarievwe, Petro~M Fochuk, Rastgo Hawrami, Anwar Hossain, et~al.
\newblock Extended defects in cdznte radiation detectors.
\newblock \emph{IEEE Transactions on Nuclear Science}, 56\penalty0 (4):\penalty0 1775--1783, 2009.

\bibitem[Roy et~al.(2019)Roy, Camarda, Cui, Gul, Yang, Zazvorka, Dedic, Franc, and James]{defects2}
Utpal~N Roy, Giuseppe~S Camarda, Yonggang Cui, Rubi Gul, Ge~Yang, Jakub Zazvorka, Vaclav Dedic, Jan Franc, and Ralph~B James.
\newblock Evaluation of cdzntese as a high-quality gamma-ray spectroscopic material with better compositional homogeneity and reduced defects.
\newblock \emph{Scientific reports}, 9\penalty0 (1):\penalty0 7303, 2019.

\bibitem[Bolotnikov et~al.(2005)Bolotnikov, Camarda, Wright, and James]{defects3}
AE~Bolotnikov, GC~Camarda, GW~Wright, and RB~James.
\newblock Factors limiting the performance of cdznte detectors.
\newblock \emph{IEEE transactions on nuclear science}, 52\penalty0 (3):\penalty0 589--598, 2005.

\bibitem[Bolotnikov et~al.(2013)Bolotnikov, Camarda, Cui, Yang, Hossain, Kim, and James]{types_defects}
AE~Bolotnikov, GS~Camarda, Y~Cui, G~Yang, A~Hossain, K~Kim, and RB~James.
\newblock Characterization and evaluation of extended defects in czt crystals for gamma-ray detectors.
\newblock \emph{Journal of crystal growth}, 379:\penalty0 46--56, 2013.

\bibitem[Ballester et~al.(2022)Ballester, Banerjee, Rodrigues, Kaspar, Vija, and Katsaggelos]{ballester2022materials}
Manuel Ballester, Srutarshi Banerjee, Miesher Rodrigues, Jaromir Kaspar, Alexander~Hans Vija, and Aggelos~K Katsaggelos.
\newblock Materials and defects characterization of cdznte sensors using the inverse synthesis method.
\newblock In \emph{2022 IEEE Nuclear Science Symposium and Medical Imaging Conference (NSS/MIC)}, pages 1--2. IEEE, 2022.

\bibitem[Banerjee et~al.(2023{\natexlab{a}})Banerjee, Rodrigues, Ballester, Vija, and Katsaggelos]{table_Nature}
Srutarshi Banerjee, Miesher Rodrigues, Manuel Ballester, Alexander~Hans Vija, and Aggelos~K Katsaggelos.
\newblock Learning-based physical models of room-temperature semiconductor detectors with reduced data.
\newblock \emph{Scientific Reports}, 13\penalty0 (1):\penalty0 168, 2023{\natexlab{a}}.

\bibitem[Banerjee et~al.(2023{\natexlab{b}})Banerjee, Rodrigues, Ballester, Vija, and Katsaggelos]{banerjee2023machine}
Srutarshi Banerjee, Miesher Rodrigues, Manuel Ballester, Alexander~Hans Vija, and Aggelos~K Katsaggelos.
\newblock Machine learning approaches in room temperature semiconductor detectors.
\newblock In \emph{X-ray Photon Processing Detectors: Space, Industrial, and Medical applications}, pages 67--94. Springer, 2023{\natexlab{b}}.

\bibitem[Banerjee et~al.(2023{\natexlab{c}})Banerjee, Rodrigues, Ballester, Vija, and Katsaggelos]{banerjee2023identifying}
Srutarshi Banerjee, Miesher Rodrigues, Manuel Ballester, Alexander~Hans Vija, and Aggelos Katsaggelos.
\newblock Identifying defects without a priori knowledge in a room-temperature semiconductor detector using physics inspired machine learning model.
\newblock \emph{Sensors}, 24\penalty0 (1):\penalty0 92, 2023{\natexlab{c}}.

\bibitem[Banerjee et~al.(2024)Banerjee, Rodrigues, Ballester, Vija, and Katsaggelos]{banerjee2024physics}
Srutarshi Banerjee, Miesher Rodrigues, Manuel Ballester, Alexander~H Vija, and Aggelos~K Katsaggelos.
\newblock A physics based machine learning model to characterize room temperature semiconductor detectors in 3d.
\newblock \emph{Scientific Reports}, 14\penalty0 (1):\penalty0 7803, 2024.

\bibitem[Ballester et~al.(2024)Ballester, Kaspar, Massanes, Banerjee, Vija, and Katsaggelos]{ballester2024modeling}
Manuel Ballester, Jaromir Kaspar, Francesc Massanes, Srutarshi Banerjee, Alexander~Hans Vija, and Aggelos~K Katsaggelos.
\newblock Modeling and simulation of charge-induced signals in photon-counting czt detectors for medical imaging applications.
\newblock \emph{arXiv preprint arXiv:2405.13168}, 2024.

\bibitem[He(2001)]{SR_review}
Zhong He.
\newblock Review of the shockley--ramo theorem and its application in semiconductor gamma-ray detectors.
\newblock \emph{Nuclear Instruments and Methods in Physics Research Section A: Accelerators, Spectrometers, Detectors and Associated Equipment}, 463\penalty0 (1-2):\penalty0 250--267, 2001.

\bibitem[Makeev et~al.(2015)Makeev, Rodrigues, Wang, and Glick]{miesher_model}
Andrey Makeev, Miesher Rodrigues, Gin-Chung Wang, and Stephen~J Glick.
\newblock Modeling czt/cdte x-ray photon-counting detectors.
\newblock In \emph{Medical Imaging 2015: Physics of Medical Imaging}, volume 9412, pages 1194--1202. SPIE, 2015.

\bibitem[Du et~al.(2003)Du, LeBlanc, Possin, Yanoff, and Bogdanovich]{Th_trapping_detrapping}
Yanfeng Du, James LeBlanc, George~E Possin, Brian~D Yanoff, and Snezana Bogdanovich.
\newblock Temporal response of czt detectors under intense irradiation.
\newblock \emph{IEEE Transactions on Nuclear Science}, 50\penalty0 (4):\penalty0 1031--1035, 2003.

\bibitem[Kamieniecki(2014)]{kamieniecki2014effect}
Emil Kamieniecki.
\newblock Effect of charge trapping on effective carrier lifetime in compound semiconductors: High resistivity cdznte.
\newblock \emph{Journal of Applied Physics}, 116\penalty0 (19), 2014.

\bibitem[Zimmerman(1973)]{zimmerman1973experimental}
W~Zimmerman.
\newblock Experimental verification of the shockley--read--hall recombination theory in silicon.
\newblock \emph{Electronics Letters}, 9\penalty0 (16):\penalty0 378--379, 1973.

\bibitem[Buttacavoli et~al.(2022)Buttacavoli, Principato, Gerardi, Cascio, Raso, Bettelli, Zappettini, Seller, Veale, and Abbene]{buttacavoli2022incomplete}
Antonino Buttacavoli, Fabio Principato, Gaetano Gerardi, Donato Cascio, Giuseppe Raso, Manuele Bettelli, Andrea Zappettini, Paul Seller, Matthew~C Veale, and Leonardo Abbene.
\newblock Incomplete charge collection at inter-pixel gap in low-and high-flux cadmium zinc telluride pixel detectors.
\newblock \emph{Sensors}, 22\penalty0 (4):\penalty0 1441, 2022.

\bibitem[Bale and Szeles(2008)]{bale2008nature}
Derek~S Bale and Csaba Szeles.
\newblock Nature of polarization in wide-bandgap semiconductor detectors under high-flux irradiation: Application to semi-insulating cd 1- x zn x te.
\newblock \emph{Physical Review B}, 77\penalty0 (3):\penalty0 035205, 2008.

\bibitem[Banerjee et~al.(2021)Banerjee, Rodrigues, Vija, and Katsaggelos]{Srutarshi_IEEE}
Srutarshi Banerjee, Miesher Rodrigues, Alexander~Hans Vija, and Aggelos~K Katsaggelos.
\newblock A learning-based physical model of charge transport in room-temperature semiconductor detectors.
\newblock \emph{IEEE Transactions on Nuclear Science}, 69\penalty0 (1):\penalty0 2--16, 2021.

\bibitem[Mont{\'e}mont et~al.(2014)Mont{\'e}mont, Lux, Monnet, Stanchina, and Verger]{diffusion_coulomb}
Guillaume Mont{\'e}mont, Silv{\`e}re Lux, Olivier Monnet, Sylvain Stanchina, and Lo{\"\i}ck Verger.
\newblock Studying spatial resolution of czt detectors using sub-pixel positioning for spect.
\newblock \emph{IEEE Transactions on Nuclear Science}, 61\penalty0 (5):\penalty0 2559--2566, 2014.

\bibitem[Benoit and Hamel(2009)]{simulation_probabilistic}
Mathieu Benoit and LA~Hamel.
\newblock Simulation of charge collection processes in semiconductor cdznte $\gamma$-ray detectors.
\newblock \emph{Nuclear Instruments and Methods in Physics Research Section A: Accelerators, Spectrometers, Detectors and Associated Equipment}, 606\penalty0 (3):\penalty0 508--516, 2009.

\bibitem[Watt et~al.(2020)Watt, Borhani, and Katsaggelos]{aggelos}
Jeremy Watt, Reza Borhani, and Aggelos~K Katsaggelos.
\newblock \emph{Machine learning refined: Foundations, algorithms, and applications}.
\newblock Cambridge University Press, 2020.

\bibitem[Sun et~al.(2019)Sun, Cao, Zhu, and Zhao]{DL_optimizers}
Shiliang Sun, Zehui Cao, Han Zhu, and Jing Zhao.
\newblock A survey of optimization methods from a machine learning perspective.
\newblock \emph{IEEE transactions on cybernetics}, 50\penalty0 (8):\penalty0 3668--3681, 2019.

\bibitem[Boyd et~al.(2004)Boyd, Boyd, and Vandenberghe]{boyd_convex}
Stephen Boyd, Stephen~P Boyd, and Lieven Vandenberghe.
\newblock \emph{Convex optimization}.
\newblock Cambridge university press, 2004.

\end{thebibliography}

\end{document}